# Evidence for two spin-glass transitions with magnetoelastic and magnetoelectric couplings in the multiferroic $(Bi_{1-x}Ba_x)(Fe_{1-x}Ti_x)O_3$ system

Arun Kumar,[1] S. D. Kaushik,[2] V. Siruguri[2] and Dhananjai Pandey[1,*]

[1]School of Materials Science and Technology, Indian Institute of Technology (Banaras Hindu University), Varanasi-221005, India.

[2]UGC-DAE Consortium for Scientific Research, Bhabha Atomic Research Centre, Mumbai 400085, India.

**ABSTRACT**

For disordered Heisenberg systems with small single ion anisotropy (D), two spin glass transitions below the long range ordered (LRO) phase transition temperature ($T_c$) has been predicted theoretically for compositions close to the percolation threshold. Experimental verification of these predictions is still controversial for conventional spin glasses. We show that multiferroic spin glass systems can provide a unique platform for verifying these theoretical predictions via a study of change in magnetoelastic and magnetoelectric couplings, obtained from an analysis of diffraction data, at the spin glass transition temperatures ($T_{SG}$). Results of macroscopic (DC M (H, T), M(t), AC susceptibility ($\chi (\omega, T)$) and specific heat ($C_p$)) and microscopic (x-ray and neutron scattering) measurements are presented on disordered $BiFeO_3$, a canonical Heisenberg system with small single ion anisotropy, which reveal appearance of two spin glass phases SG1 and SG2 in coexistence with the LRO phase below the *A-T and G-T lines*. It is shown that the temperature dependence of the integrated intensity of the antiferromagnetic (AFM) peak shows dips with respect to the Brillouin function behaviour around the SG1 and SG2 transition temperatures. The temperature dependence of the unit cell volume departs from the Debye-Grüneisen behaviour below the SG1 transition and the magnitude of departure increases significantly with decreasing temperature upto the electromagnon driven transition temperature below which a small change of slope occurs followed by another similar change of slope at the SG2 transition temperature. The ferroelectric polarisation also changes significantly at the two spin glass transition temperatures. These results, obtained using microscopic techniques, clearly demonstrate that the SG1 and SG2 transitions occur on the same magnetic sublattice and are intrinsic to the system. *We also construct a phase diagram showing all the magnetic phases in BF-xBT system. While our results on the two spin glass transitions support the theoretical predictions, it also raises several open questions which need to be addressed by revisiting the existing theories of spin glass transitions by taking into account the effect of magnetoelastic and magnetoelectric couplings as well as electromagnons.*

# I. Introduction:

Study of phase transitions in disordered magnetic systems has been a time honoured problem in the field of solid state and materials sciences. In dilute magnetic systems (e.g. Cu-Mn), the ground state is known to be a spin glass (SG) state [1-4]. However, controversies still abound in the case of concentrated systems. Theoretically, it is known that the disordered concentrated magnetic systems can still lock into a long range ordered (LRO) magnetic ground state if the disorder content (c) is less than a percolation threshold ($c_p$) for the exchange pathways, except that there is disorder induced broadening of the phase transition leading to the rounding of the susceptibility peak at the transition temperature ($T_c$) [1,5,6]. However, when the disorder content is close to the percolation threshold, the LRO percolative phase for both the Ising [7-11] and Heisenberg [12-16] systems has been reported to undergo another transition to the SG state. The pertinent questions that arise in relation to such systems are: (1) What is the true ground state? (2) Does the LRO phase coexist with SG phase in the ground state? (3) If both the phases do coexist, what is the proof that the SG phase has resulted from the same magnetic sublattice that led to the LRO phase? (4) Is the coexistence of SG phase with the LRO phase due to coexistence of isolated short range ordered (SRO) superparamagnetic (SPM) clusters with LRO clusters on two different magnetic sublattices as a result of segregation and clustering?

The theoretical treatments for such concentrated Ising as well as Heisenberg systems predict that the SG state can result from freezing of either the longitudinal or transverse components of the spin in the LRO phase and that it can coexist with the LRO phase on the same magnetic sublattice [17-26]. These theoretical predictions cannot be verified using macroscopic measurements (DC and AC susceptibilities) alone and require microscopic tools (neutron scattering, Mössbauer spectroscopy etc.) which have been used for a few systems in support of the longitudinal/transverse freezing model in both the Ising and Heisenberg systems [7-16]. More interestingly, yet another interesting situation has been predicted theoretically for concentrated Heisenberg systems with small single ion anisotropy (D) as compared to the magnetic exchange interaction (J), where both the longitudinal and transverse components can freeze successively leading to two SG transitions below the so-called Almeida-Thouless (A-T) and Gabey-Toulose (G-T) lines, respectively [23-26]. Although most of these theoretical treatments are for concentrated ferromagnetic (FM) systems, these theories have been applied to disordered antiferromagnetic (AFM) systems as well [7]. While evidence for two SG transitions has been obtained in several systems using macroscopic measurements [27-34], the results are rather controversial as it is not clear if the LRO and the SG phases are formed on the same or separate magnetic sublattices. Unambiguous evidence confirming the occurrence of two SG transitions and coexistence of the two SG phases (SG1 and SG2) with the LRO phase on the same magnetic lattice using microscopic tools are rather sparse [e.g. Ref. 29] in such systems.

Spin glass phase has been reported in several multiferroic systems also [35-40]. Unlike the conventional SG systems, the multiferroic SG systems offer the possibility of verifying the theoretical predictions for concentrated systems through a study of the change in ferroelectric polarization and strain as a result of magnetoelectric and magnetoelastic couplings across the SG transition using diffraction techniques. A transition from LRO AFM phase to SG phase at low temperatures with coexistence of LRO and SG phases in the ground state has been reported in the multiferroic systems like pure [35-37] and disordered [38] BiFeO$_3$ and some site-disordered compounds like Pb(Fe$_{1/2}$Nb$_{1/2}$)O$_3$ [39,40]. The origin of coexistence of SG and LRO AFM phases at low temperatures in such multiferroics is still controversial as experimental evidences for and against both the phase segregation [39] and transverse freezing models [40] have been advanced in the literature. Further, there is no experimental report for LRO to LRO+SG1 to LRO+SG2 transitions in a disordered multiferroic systems. In this context, it is interesting to note that the single ion anisotropy (D) of $BiFeO_3$, a canonical Heisenberg system, is rather small as compared to exchange interaction (J) D/J ~0.001 [41-44] which falls within the range where two SG phases have been reported in non-multiferroic disordered systems like MgMn [24].

Here we present first experimental evidence in support of theoretical predictions for *two SG phases below the A-T and G-T lines in coexistence with the LRO phase* on the same magnetic sublattice in the multiferroic solid solution $(Bi_{1-x}Ba_x)(Fe_{1-x}Ti_x)O_3$ (BF-xBT) system using a combination of macroscopic (DC M(H,T), M(t), AC susceptibility ($\chi$ ($\omega$, T)) and specific heat ($C_p$)) and microscopic (x-ray and neutron scattering) measurements. We have selected $BaTiO_3$ based solid solution of $BiFeO_3$ for this study as it has received considerable attention in recent years due to large ferroelectric polarization [45-47], highest depolarization temperature for piezoelectric applications [47,48] and destruction of spin cycloid [45,46,49-52] leading to large remnant magnetization [45,46,49-52] as well linear magnetoelectric coupling [45,46]. From the analysis of neutron and x-ray diffraction data on BF-0.20BT, we demonstrate two distinguishing features of SG transitions in disordered multiferroics: (1) very strong and moderate magnetoelastic couplings associated with the SG1 and SG2 transitions, respectively, as revealed by the change in the unit cell volume ($\Delta V$) with respect to the theoretically predicted values, that scales quadratically with the spontaneous magnetization ($M_s^2$) and (2) strong magnetoelectric coupling at both the SG transitions as revealed by the large change in spontaneous polarization ($\Delta P_s$), calculated from the atomic coordinates obtained by Rietveld refinements of the nuclear structure and the Born effective charges (BEC), at the two SG transitions. After presenting the results on BF-0.20BT, we also discuss the effect of dopant ($BaTiO_3$) concentration on the magnetic transitions in BF-xBT and construct a phase diagram showing all the magnetic phases.

## II. Experimental:

### (a) Sample preparation:

Polycrystalline samples of $(Bi_{1-x}Ba_x)(Fe_{1-x}Ti_x)O_3$ (BF-xBT) solid solutions were synthesized by solid state route for x = 0.10 to 0.60 at Δx = 0.10 interval using high purity oxides of $Bi_2O_3$, $Fe_2O_3$, $BaCO_3$, $TiO_2$, $MnO_2$ (Aldrich and Alfa Aesar). The starting materials were carefully weighed in stoichiometric ratio and mixed in an agate mortar and pestle for 3 hours and then ball milled for 6 hours in acetone as mixing media using zirconia jar and zirconia ball. After drying, the mixture was calcined at optimized temperatures in the range 1073 K-1173 K depending upon the composition for 8 hours in open alumina crucible. The calcined powders were mixed with 0.3 wt % $MnO_2$ and ball milled again for 4 hours to break the agglomerates formed during calcination. $MnO_2$ doping reduces the leakage current as discussed in the literature [47]. We used 2 % polyvinyl alcohol as a binder to press the calcined powder into pellets of 12 mm diameter and 1 mm thickness at an optimized load of ~70 kN. After the binder burn-off at 773 K for 12 hours, sintering were carried out at optimized temperatures in the range 1173 to 1273 K, in closed alumina crucible with calcined powder of the same composition as spacer powder for preventing the loss of $Bi_2O_3$ during sintering. The sintering time was increased with increasing $BaTiO_3$ content from 1 hour for x = 0.10 to 4 hours for x = 0.60.

### (b) Experimental details:

X-ray powder diffraction (XRD) measurements in the temperature range 12 K to 350 K were carried out using an 18-kW Cu rotating anode powder diffractometer (Rigaku) operating in the Bragg-Brentano geometry and fitted with a curved crystal monochromator in the diffraction beam. Sample environment was varied using a close cycle helium refrigerator based low temperature attachment on this diffractometer. The data in the 2θ range 20 to 120° were collected using annealed powders (10 hours at 773 K) obtained after crushing the sintered pellets at a step of 0.02 degrees. High resolution synchrotron x-ray powder diffraction (SXRD) patterns were also recorded at PETRA III, Germany at 60 keV energy for a few selected temperatures above liquid $N_2$ temperature. Temperature dependent neutron powder diffraction (NPD) data in the range 300 K to 2.8 K was collected at Druva reactor, BARC, Mumbai at a wavelength of 1.48 Å using high-resolution powder diffractometer. Composition analysis was carried out using Electron Probe Micro Analyzer (EPMA) and CAMECA SXFive instrument. The nuclear and magnetic structures were refined by Rietveld techniques using FULLPROF suite [53]. DC magnetization (M(T, H)) measurements were carried out on a SQUID based magnetometer (Quantum Design, MPMS-3) in the temperature range 2 K to 900 K at 500 Oe applied dc field in two separate measurements from 2 to 400K and 300K to 900K range. The ac susceptibility ($\chi(\omega, T)$) measurements were carried out in the temperature range 2 K to 300 K on the same machine using an ac drive field of 2 Oe. The heat capacity ($C_p$) measurement was carried out in the temperature range 1.8 to 387 K using physical property measurement system (PPMS) (Dynacool, Quantum Design, USA).

**III. Results and Discussion:**

**A. Magnetic transitions in BF-0.20BT:**

The antiferromagnetic (AFM) transition in pure $BiFeO_3$ (BF) occurs at $T_N$ ~643K. As a result of 20% substitution of $BaTiO_3$ in $BiFeO_3$, i.e. in BF-0.20BT, $T_N$ decreases due to dilution of the magnetic sublattice. Fig.1 depicts the zero-field cooled (ZFC) DC magnetization (M(T)) at an applied field of 500 Oe in the temperature range 2-900K. It is evident from the figure that a long range ordered (LRO) magnetic phase emerges below $T_N$ ~608K in agreement with the previous results [49]. The nature of ZFC M(T) response of BF-0.20BT is, however, not like a typical AFM transition seen in pure BF but is rather like a ferromagnetic (FM) transition. The FM type transition is due to the destruction of the spin cycloid, superimposed on the canted G-type AFM arrangement of spins in BF, that releases the latent FM component of the spins in magnetic sublattice. This was confirmed through M-H hysteresis loop measurements, Curie-Weiss plot and neutron diffraction patterns.

The M-H hysteresis loop at 300 K for BF-0.20BT reveals weakly ferromagnetic behaviour (see Fig. 2) in contrast to linear M-H characteristic of AFM phase in pure BF. However, even in pure BF, the M-H loop opens up with a remanant magnetization $M_r$ ~ 0.3 emu/g at 10 K on destruction of the spin cycloid in the presence of external magnetic field in excess of ~18T [54,55]. The opening of the hysteresis loop in BF-0.2BT even at moderate fields thus indicates the destruction of the spin cycloid of $BiFeO_3$ as noted by previous workers also in various solid solutions of BF [49-52]. The remanant magnetization $M_r \approx 0.13$ emu/g of our samples is close to the value of ~0.15 emu/g reported by Singh et al. [49]. The fact that the magnetization does not saturate even at 7T field also suggests weakly FM behaviour due to canted AFM structure.

The ZFC M (T) of BF-0.20BT shows Curie-Weiss behaviour $\chi = C/(T-\theta_W)$, where C and $\theta_W$ are Curie constant and Curie-Weiss temperature, respectively. Fig.1(b) shows the temperature dependence of inverse DC susceptibility ($\chi^{-1}$) whose linear behaviour at high temperatures (T > 700 K) clearly confirms to Curie-Weiss law with $\theta_W$ = -873.6 K. The large negative value of $\theta_W$ indicates strong antiferromagnetic interactions in the LRO AFM state. The effective magnetic moment ($\mu_{eff}$) of $Fe^{3+}$ ion, calculated from the Curie constant C, comes out to be 4.98 $\mu_B$ which is nearly 80% of the magnetic moment of $Fe^{3+}$ ions in the high spin configuration (S = 5/2) as expected for BF-0.20BT due to 20% Ti substitution at the Fe site.

AFM structure of BF-0.20BT was further confirmed by neutron powder diffraction (NPD) studies. Fig.3 shows the NPD pattern of BF-0.20BT at room temperature in the limited 2θ range of 15º -57º. This pattern contains main perovskite reflections as well as some superlattice reflections which arise either due to antiferromagnetic ordering or tilting of oxygen octahedra. All the reflections could be indexed with respect to a doubled perovskite unit cell. The $111_{pc}$ (pc stands for pseudocubic unit cell) magnetic superlattice peak at 2θ = 18.6º (marked with an arrow) is not allowed in the rhombohedral R3c space group and arises due to AFM ordering of the Fe spins. Thus, the transition at $T_N$ ~608K in Fig.1 is linked with the appearance of a long range ordered (LRO) AFM phase.

Below room temperature, the ZFC M(T) of BF-0.20BT clearly reveals three anomalies near 240K, 140K and 30K (see inset (a) of Fig. 1). In addition, the ZFC and FC M(T) curves show bifurcation due to history dependent effects. Such bifurcation has been reported in spin glass and superparamagnetic (SPM) systems [1-4,56]. In canonical spin glasses, ZFC M(T) shows a cusp at $T_{max}$ and the bifurcation of FC and ZFC M(T) occurs close to the cusp temperature [1-4]. However, unlike the canonical systems, the peak around ~240 K in M(T) of BF-0.2BT is quite smeared out and the bifurcation starts well above $T_{max}$. While smeared peak have been reported in several in cluster glass and SPM systems due to occurrence of freezing/blocking over a wide range of temperatures as a result of large distribution of cluster sizes [57-59], the peak around 240K is much more broad and the bifurcation of ZFC and FC M (T) curves occurs well above the peak temperature ($T_{max}$). The extent of broadening of the 240K peak in the ZFC M(T) measurements is dependent on the field strength as discussed in section

C). As the specific heat can probe any magnetic transition with higher sensitivity than the magnetization measurements, we carried out specific heat measurements also. Fig.4 depicts the variation of specific heat ($C_p$) with temperature which reveals a weak but much sharper anomaly (see inset (a)) corresponding to the 240K transition in ZFC M(T). As shown in section B, the AC susceptibility peak is also relatively sharper (see inset of Fig.5(b)) than the peak in the ZFC M(T) for the 240K transition. Obviously, the time scales associated with different measuring probes give different widths for 240K transition as expected for a glassy phase in a concentrated system with larger distribution of cluster sizes. What is significant is that all the three different measurement probes, i.e. M(T), AC susceptibility, and specific heat, clearly confirm that a transition is indeed taking place around 240K.

Below the 240K transition, the ZFC M(T) plot shows a kink around 140 K followed by a nearly temperature independent plateau upto ~30K. On further cooling below 30K, ZFC M(T) starts decreasing. The FC M(T) also shows a kink around 140K but below this temperature it keeps on increasing without any anomalous decrease around 30K. In polycrystalline $BiFeO_3$ sample [37] and single crystals of $BiFeO_3$ [35], two transitions around 250K and 30K, respectively have been reported but not in the same sample. *The transition around 140K has been investigated in great detail in $BiFeO_3$ and has been linked with electromagnons [60-64]. The electromagnons are collective spin and lattice excitations and can be excited by electric field. The electromagnons have been reported by terahertz [65] and Raman spectroscopies [60-62] as well as inelastic neutron scattering studies [41,42]. The first experimental evidence of electromagnons was demonstrated in $RMnO_3$ (R = Tb, Gd) using terahertz spectroscopy [65] whereas in $BiFeO_3$, the electromagnons were first reported using Raman spectroscopy [60-62] where the intensity and frequency of magnon modes appearing* *around 140K were shown to change on application of external electric fields. The theoretical work of de Sousa and Moore [64] and Fishman et al. [42, 44] have confirmed the existence of electromagnons in Raman scattering studies on BiFeO₃. In case of BF-0.20BT, the M(T) measurement reveals strong signature of 140K (±5K) transition and shows an anomaly in the integrated intensity of the AFM peak in the neutron diffraction pattern (discussed later in section E). We believe that this transition is also linked with electromagnons although, Raman scattering, THz spectroscopy and inelastic neutron scattering studies are required to confirm this. As this is beyond the scope of the present work, we keep our focus on the other two transitions occurring around 240K and 30K in what follows hereafter.*

## B. Evidence for two spin glass transitions in BF-0.20BT:

We carried out frequency dependent AC magnetic susceptibility ($\chi(\omega, T)$) measurements to understand whether the bifurcation of the ZFC and FC M(T) is associated with spin glass freezing or SPM blocking. Figs. 5 (a) and (b) depict real ($\chi'(\omega, T)$) and imaginary ($\chi''(\omega, T)$) parts, respectively, of $\chi(\omega, T)$ of BF-0.20BT measured at various frequencies for a drive field of 2 Oe in the temperature range 2-300K. The $\chi'(\omega, T)$ shows two peaks at $T_{f1}$ and $T_{f2}$ corresponding to the two anomalies around ~240 and ~30K revealed in ZFC M(T) plot as can be seen from the insets (i) and (ii) of Fig. 5(a). It is noteworthy that the temperature dependence of $\chi''(\omega, T)$ for the 240K anomaly exhibits normal freezing behavior whereas it shows anomalous behavior with negative cusp for the 30K anomaly. The negative cusp is in agreement with that reported in single crystals of BiFeO₃ as well as in polycrystalline samples of BiFeO₃ [35-36]. The anomalous frequency dependence of the lowest temperature SG phase (SG2) has been discussed in detail in the context of pure BiFeO₃ where the role of cycloidal magnetic structure has been highlighted [35]. However, the spin cycloid of BiFeO₃ is known to be destroyed in the presence of disorder, such as 20% BaTiO$_3$ substitution in the present case. This has been confirmed by neutron scattering and magnetization measurements [46,49]. Suffice is to say that the opening of the M-H loop in our samples (see Fig. 2) rules out the presence of spin cycloid and therefore there is no correlation between the anomalous frequency dispersion [see Ref. 35 for more details] of the 30K anomaly and the spin cycloid.

*The peak corresponding to the 240K anomaly in ZFC M(T) plot is relatively less broad in χ' (ω, T) and χ'' (ω, T) as compared to that in the ZFC M(T) indicating the role of time scales associated with the spin freezing/blocking process and the measurement time for different probes.* The temperatures $T_{f1}$ and $T_{f2}$ corresponding to the two peaks in χ' (ω, T) shift towards higher side on increasing the measuring frequency. Such a frequency dependent shift may be due to either SG freezing or SPM blocking [1-4,56]. The shift of the χ' (ω, T) peak temperature has been analyzed in terms of an empirical frequency sensitivity parameter $K = \Delta T_f / (T_f \Delta(\ln \omega))$ (the so-called Mydosh parameter) which lies in the range 0.003-0.08 [66-68] and 0.1 to 0.3 [66] for spin-glass freezing and SPM blocking, respectively. In the case of BF-0.20BT, K is found to be ~0.04 for both the transitions which supports the spin glass freezing rather than SPM blocking.

For SPM blocking, the relaxation time (τ) should follow the typical Arrhenius type dependence without any critical behaviour [56]:

$$\tau = \tau_0 \exp(E_a / k_B T), \quad (1)$$

where τ is the relaxation time, $E_a$ the activation energy, $k_B$ the Boltzmann constant, and $\tau_0$ the inverse of the attempt frequency. The ln τ vs 1/T plots derived from the frequency dependent peak positions $T_f(\omega)$ of χ' (ω, T) for the transitions around 240 K and 30 K are therefore expected to be linear for SPM blocking. The fact that this plot is non-linear in BF-0.20BT, as can be seen from Figs. 6(a) and (b)), rules out the SPM blocking being responsible for the two peaks in $\chi'$ ($\omega$, T).

For spin glass freezing, one observes critical slowing down of the relaxation time ($\tau$) due to ergodicity breaking. This has been modeled using a power law [69-70]:

$$\tau = \tau_0[(T_f - T_{SG})/T_{SG}]^{-z\nu}, \qquad (2)$$

where, $T_{SG}$ is the SG transition temperature, $\nu$ the critical exponent for the correlation length ($\xi$) and z the dynamical exponent relating $\tau$ to $\xi$. In some spin glass systems [71], the frequency dependent shift of the $\chi'$ ($\omega$, T) peak temperature has been modeled using the empirical Vogel–Fulcher (V-F) law also :

$$\tau = \tau_0 \exp(E_a/k_B (T-T_{SG})), \qquad (3)$$

where $E_a$ is the activation energy. Both the power law and V-F law type critical dynamics provide excellent fits for the two transitions as can be seen from Figs. 7(a) and (b), respectively. The fitting parameters for the two transitions are: $T_{SG1} \sim$ (218.6±0.8) K, $z\nu_1$ = 2.09 s, $\tau_{01}$ = 3.87x10$^{-6}$ s and $T_{SG2}$ = (18.6±0.4) K, $z\nu_2$ = 0.69, and $\tau_{02}$ = 1.92x10$^{-4}$s for power law and $T_{SG1}$ ~(214±2) K, $E_{a1}$ = 4.89 meV, and $\tau_{01}$ = 5.64x10$^{-6}$s and $T_{SG2}$ ~(15.9±0.1) K, $E_{a2}$ = 0.65 meV, and $\tau_{02}$ =1.64x10$^{-4}$ s for V-F law. The continuous line in Figs. 6(a) and (b) are the fits using these parameters in the ln $\tau$ vs 1/T plots. Both the fits are excellent. The values of $T_{SG1}$ and $T_{SG2}$ as well as $\tau_{01}$ and $\tau_{02}$ obtained by V-F law and power law type critical dynamics are comparable. Thus, both the power law and V-F dynamics confirm the glassy nature of the two frequency dependent anomalies in $\chi'$ ($\omega$, T). The magnitude of $\tau_{01}$ and $\tau_{02}$ for both the power law and V-F law type dynamics falls in the typical cluster glass (CG) category (10$^{-5}$-10$^{-10}$s) for concentrated systems [1,66] and not the canonical spin glasses in dilute systems [1].

**C. Evidence for de Almeida-Thouless and Gabay-Toulouse lines in BF-0.20BT:**

*The existence of two spin glass phases, which we shall label as SG1 and SG2 hereafter, was further confirmed by the presence of the so-called de Almeida-Thouless (A-T) [18, 23-25] and Gabay-Toulouse (G-T) [23-25] lines. For Ising systems, it has been shown by de Almeida and Thouless [18] that the peak temperature ($T_{max}$) of the ZFC M (T) plot shifts to lower temperature side on increasing the magnetic field (H) as a result of replica symmetry breaking [18]. For low fields, this shift shows the following H dependence:*

$$H^2 = A \, [1 - T_{max}(H)/T(0)]^3, \qquad (4)$$

*where $T_{max}(H)$ and $T(0)$ are the field dependent and zero-field freezing temperatures, respectively. Eq. (4) sets the boundary between the ergodic paramagnetic and non-ergodic spin glass phases and is commonly known as the A-T line [18]. For the Heisenberg systems also, it has been shown that the A-T line is present and $T_{max}$ follows $H^{2/3}$ dependence at low fields [24-25]. However, it can occur due to freezing of either the longitudinal ($q_{ll}$) or the transverse ($q_\perp$) components of the spin, depending on whether the single ion anisotropy (D/J) is positive or negative. For low values of D/J, a second SG transition whose $T_{max}$ decreases as $H^2$ at low fields is predicted to occur due to the freezing of the second component of the spin. For small but positive values of D/J, as is the case with $BiFeO_3$ [41-44], the first SG transition (i.e. SG1) is expected to be due to the freezing of $q_{ll}$ component while the second one (i.e. SG2) due to freezing of $q_\perp$ as per the theoretical predictions [24-25]. The H dependence of the $q_{ll}$ and $q_\perp$ freezing temperatures should thus fix the A-T and G-T lines in the $T_{max}$ versus H phase diagram for the SG1 and SG2 phases, respectively..*

*To verify the existence of A-T and G-T lines in BF-0.20BT, we carried out ZFC M (T) measurements at different fields and the results are depicted in Fig. 8 for both the transitions. It*

*is evident from the figure that the peak corresponding to SG1 transition is prominent, even though broad, while no such peak is observed for SG2 transition up to a field of 500 Oe. With increasing field, the peak corresponding to SG2 transition also starts taking a prominent shape (see insets) while the peak corresponding to the SG1 transition starts getting smeared and suppressed after initial sharpening upto 800 Oe. We find that the $T_{max}$ for both the transitions decreases with increasing magnetic field as expected theoretically. The linear nature of the $T_{max}$ versus $H^{2/3}$ and $T_{max}$ vs $H^2$ plots shown in Figs. 9(a) and 9(b) for the SG1 and SG2 transitions confirms the existence of A-T and G-T lines, respectively, in the $T_{max}$ versus H phase diagram. Thus our results confirm the theoretical predictions [23-25] for two spin glass transitions in Heisenberg systems with low D/J.*

**D. Relaxation of thermoremanent magnetization for the spin glass phases of BF-0.20BT:**

Spin glass state is known to exhibit slow relaxation of thermoremanent magnetization which has been modelled using stretched exponential function [67,72,73]:

$$M(t) = M_0 + M_r \exp[-(t/\tau)^{1-n}] \qquad (5)$$

where $M_0$ is the intrinsic static magnetization component, $M_r$ the glassy component, $\tau$ the characteristic relaxation time and n the stretched exponential exponent. To study the slow relaxation of the thermoremanent magnetization, we cooled the sample under a field of 1T from 300 K to 200K for the SG1 phase. After reaching the set temperatures, the sample was allowed to age without switching off the field for a waiting time of $t_w$ = 500s. After the elapse of the waiting time $t_w$, the field was switched off. For the SG2 phase, the sample was first annealed at 773 K above $T_N$ to remove any remanent magnetization introduced during the first cycle and then cooled to 10K under 1T field. This was followed by the protocol identical to that adopted for the SG1 phase. The thermoremanent magnetization so measured as a function of time is shown in Figs. 10 (a) and (b) at 200K and 10K, respectively. The continuous line in the two figures depicts the best fit for Eq. (5). These fits yield n, $M_0$, $M_r$ and $\tau$ as 0.55, 0.1575 emu/g, 0.0008 emu/g, (1207±15)s for the SG1 phase and 0.53, 0.1697 emu/g, 0.0009 emu/g, (1661±14)s for the SG2 phase, respectively. The observed exponent (n) and relaxation time ($\tau$) are in agreement with the reported values for cluster glasses and super spin glasses [67,73]. Thus, relaxation behaviour of thermoremanent magnetization also favours the existence of two SG phases in BF-0.20BT.

**E. Evidence for magnetoelastic coupling at spin glass transitions in BF-0.20BT:**

In order to verify if the two SG transitions and the intervening transition driven by electromagnons involve any structural phase transition, we carried out XRD studies in the temperature range 12K to 350K. Fig.11 depicts the temperature evolution of the XRD profiles of a few selected pseudocubic (pc) peaks ($222_{pc}$, $400_{pc}$ and $440_{pc}$ reflections) of BF-0.20BT after stripping off the $K_{\alpha 2}$ contribution. It is evident from this figure that the $222_{pc}$ and $440_{pc}$ peaks are doublets, whereas $400_{pc}$ is a singlet, as expected for the rhombohedral structure, down to 12K which implies absence of any structural phase transition below room temperature. This was further confirmed by Rietveld refinements at different temperatures. It was found that the rhombohedral R3c space group gives excellent fit between the observed and calculated profiles at all temperatures down to 12 K. The details of the refinement are presented in section S3 of the supplemental information.

While the magnetic measurements clearly indicate the existence of SG1 and SG2 transitions in BF-0.20BT, the reason for the broad nature of the peak in the ZFC M(T) of the SG1 transition needs to be understood. *In order to rule out the role of a structural phase transition, which might have been missed in the medium resolution rotating anode based XRD data, we also carried out Rietveld refinement using high resolution synchrotron x-ray diffraction (SXRD) patterns at three selected temperatures 260K, 240K and 220K. Fig.12 depicts the observed, calculated and difference profiles obtained after the Rietveld analysis of the SXRD patterns at 260K, 240K and 220K, respectively, for BF-0.20BT using R3c space group. The excellent fit between observed and calculated profiles confirms that the R3c space group for BF-0.2BT at room temperature does not change across the SG1 transition. We can thus conclusively rule out the role of any structural phase in the broad SG1 transition.*

Even though there is no structural phase transition, the temperature dependence of unit cell volume ($V_{hex}$), as obtained from the Rietveld refinements, shows anomalies around the three magnetic transitions (see Fig. 13). It is interesting to note that the slope of the experimental $V_{hex}$ versus T plot changes prominently around the SG1 transition without any discontinuous change in the value of $V_{hex}$. After the initial change of slope, the experimental $V_{hex}$ values decrease smoothly with temperature below SG1 transition upto ~150K. Small changes in volume around 140 and SG2 transitions are also observed as shown in the inset (a) of the figure. The large change of slope around the SG1 transition suggests strong magnetoelastic coupling associated with this transition. It is possible to separate out the magnetic (magnetoelastic) contribution from the anharmonic lattice part at least for the SG1 transition because of the large slope change. For this, the temperature dependence of $V_{hex}$ above $T_{SG1}$ was modeled using the Debye-Grüneisen equation:

$$V \cong V(0) + \frac{9\gamma N k_B}{B} T \left(\frac{T}{\Theta_D}\right)^3 \int_0^{\Theta_D/T} \frac{x^3}{e^x - 1} dx \qquad (6),$$

where V(0), $\Theta_D$, $\gamma$ and B are the unit cell volume at 0K, the Debye temperature, the Grüneisen parameter and the bulk modulus, respectively. Continuous solid line in the figure shows the results of least squares fit to the observed unit cell volume in the temperature range 260K <T≤ 350K using Eq. (6). The fitting parameters so obtained are: V (0) = (375.86 ± 0.01) Å$^3$, $\Theta_D$ = (494 ± 39) K, and $9\gamma Nk_B/B$ = (0.071 ± 0.003) Å$^3$/K. The difference $\Delta V$ between the experimentally observed values of $V_{hex}$ and the theoretically calculated anharmonic lattice contribution increases with decreasing temperature. It is interesting to note that the bulk strain ($\Delta V/V$) vs $M_s^2$ plot corresponding to the shaded region in the figure is linear in the temperature range 240 to 150K as can be seen from inset (b) of Fig. 13. This linear dependence confirms that the slope change is due to quadratic spin-lattice coupling [74]. *The fact that the change of slope is much more pronounced around SG1 as compared to that around 140 and SG2 transitions suggests that the spin-lattice coupling for the other two transitions is rather weak as compared to that for the SG1 phase.*

### F. Evidence for coexistence of LRO AFM and spin glass phases in BF-0.20BT

We now turn towards neutron diffraction studies to understand whether the LRO, SG1 and SG2 transitions occur on the same magnetic sublattice or not. Fig.14 depicts the temperature evolution of the neutron powder diffraction patterns of BF-0.20BT in the limited 2θ range of 15-57º. It was verified by Rietveld refinement that neither the nuclear nor the magnetic structure changes down to the lowest temperature of measurement (see section S4 of the supplemental information for more details). The fact that the AFM peak, marked with arrow in the figure, persists down to 2.8K clearly suggests that the LRO AFM phase coexists with the SG phases. We modelled the temperature dependence of the integrated intensity of the AFM peak using the molecular-field theory according to which the magnetic moment should follow the following temperature dependence [75],

$$\frac{\mu}{\mu_0} = B_J(x) \text{ where, } x = \left(\frac{3J}{J+1} \frac{T_C}{T} \frac{\mu}{\mu_0}\right) \quad (7)$$

where J is the total angular momentum of the system, $\mu/\mu_0$ is the ratio of the magnetic moment at temperature T to that at T= 0K, and $B_J$ is the Brillouin function

$$B_J(x) = \frac{2J+1}{2J}\coth\left(\frac{2J+1}{2J}x\right) - \frac{1}{2J}\coth\left(\frac{1}{2J}x\right) \quad (8)$$

We fitted the square of the ordered magnetic moment to the experimentally measured integrated intensity of the AFM peak as a function of temperature and the results are shown in Fig.15. Solid line in the figure is the fit for the square of the Brillouin function behaviour. Evidently, the observed variation of the integrated intensity of the AFM peak deviates from the mean field behavior around the two SG transition temperatures. This decrease in the integrated intensity around $T_{SG1}$ and $T_{SG2}$ clearly suggests that some spin/spin components are being removed from the LRO AFM phase regions and transformed to the glassy phase. This proves that the two SG phases are formed on the same magnetic sublattice [40] that gives rise to the LRO AFM phase and that they are not due to nanosized impurity phases, proposed in the context of the low temperature SG phase of pure $BiFeO_3$ [76-78] or smaller SPM clusters in a segregated magnetic microstructure proposed in the context of $Pb(Fe_{1/2}Nb_{1/2})O_3$ [39].

**G. Evidence for isostructural phase transitions and polarisation changes across spin glass transitions in BF-0.20BT:**

*Even though the space group symmetry of BF-0.20BT does not change in the 300 to 2.8K temperature range, the fractional coordinates of $z_{Bi/Ba}$ and $z_{Fe/Ti}$, obtained by Rietveld*

*refinements using neutron diffraction data, change discontinuously around the two spin glass transition temperatures as shown in Fig.16. Further, the coordinates of the two oxygen positions ($x_O$ and $y_O$) show anomalies around the third transition driven by electromagnons. This change of atomic positions (fractional coordinates) can be explained in terms of one of the irreducible representations (Irrep) of the R3c space group corresponding to an optical phonon mode at k= 0,0,0 point of the Brillouin zone, as discussed in the supplemental information of Ref. 45. Such a change of atomic positions without any change in the space group symmetry has previously been observed in BF solid solutions across $T_N$ where it has been attributed to an isostructural phase transition (ISPT) [45,46,79]. We believe that the anomalies in atomic positions across the three low temperature magnetic transitions in BF-0.20BT are due to similar ISPTs driven by spin-polar phonon coupling (SPC). In literature [80], the origin of SPC effect has been attributed to the electronic structure which may suggest that the low temperature transitions in BF-0.20BT could be of electronic origin. However, the calculations also indicate that the electronic contributions to the SPC effect in BF is rather small [80].*

As a result of change in the atomic positions due to the ISPT, the ferroelectric polarisation ($P_s$) is known to change significantly by about 2 to 3 µC/cm² at $T_N$ revealing magnetoelectric coupling in BiFeO$_3$ solid solutions including BF-0.20BT [45,46,79]. We have also calculated $P_s$ below room temperature from Rietveld refined coordinates, unit cell parameters and first principles derived Born Effective Charges (BEC) taken from the literature [81] using the following relationship:

$$P = e/V \sum_k z'_k \Delta(k), \qquad (9)$$

where the sum runs over all the ions inside the unit cell while $\Delta(k)$ is the displacement of the k$^{th}$ ion from its ideal cubic perovskite position, $z'_k$ the Born effective charge for k$^{th}$ ion and V the volume of the primitive unit cell. The temperature variation of $P_s$ so obtained is shown in Fig.17 which reveals distinct changes across the two SG transitions. The change in $P_s$ observed by us around $T_{SG1}$ and $T_{SG2}$ is $(5\pm1)\mu C/cm^2$ and $(2\pm1)\mu C/cm^2$, respectively, which are of similar order of magnitude as reported at the $T_N$ for BF-0.20BT [46]. The observation of change in $P_s$ ($\Delta P_s$) at the two SG transitions not only reveals strong magnetoelectric coupling but also provides additional microscopic evidence for the coexistence of the SG and the LRO phases on the same magnetic sublattice at the two spin glass transition temperatures due to multiferroic nature of the two SG phases.

**H. Magnetic phase diagram of BF-xBT:**

*Before we conclude, we would like to discuss the effect of BT concentration (x) on the low temperature phase transitions in BF with the objective of constructing a magnetic phase diagram of BF-xBT system using the transition/freezing temperatures obtained from ZFC M(T) and AC susceptibility measurements (see Figs 18- 20). Fig. 18 depicts the plot of ZFC M(T) for various compositions (x). Signature of a transition to a LRO magnetic state is clearly seen upto x=0.40. For x=0.50 also, a diffuse transition is seen in the figure but for x=0.60 there is no signature of this transition in the M(T) plot. Disorder induced gradual broadening of the transition is seen quite clearly in this figure for high x values. The LRO transition temperature $T_N$ was determined from the first derivative of M(T) which shows clear dips for to all the compositions including x=0.50 (see also inset of Fig.18). The composition dependence of $T_N$ shown in Fig. 21 could be described using $(x-x_c)^n$ type dependence with n =0.30±0.02 and $x_c$ = 0.55±0.01. In the previous neutron diffraction studies [49], AFM peak was observed for x=0.50 but not for x=0.60 which also suggests that $x_c$ lies in the range $0.50 \leq x_c \leq 0.60$. We believe that*

$x_c$= 0.55 is the percolation threshold limit for the LRO phase to emerge in the presence of disorder introduced by BT substitution in the BF matrix.

To investigate the effect of disorder (x) on the SG1 and SG2 transitions, we show in *Fig. 19* the $\chi'(\omega, T)$ plots at 497.3 Hz for various compositions of BF-xBT. The variation of $\chi'(\omega, T)$ for x = 0.10, 0.20 and 0.30 are similar where the peaks corresponding to SG1 and SG2 transitions are clearly seen. While two peaks in the $\chi'(\omega, T)$ plot are also seen for x= 0.40, the magnitude of the susceptibility below the SG1 transition shows a slightly increasing trend with decreasing temperature whereas it shows a decreasing trend for x = 0.10, 0.20 and 0.30 showing that the disorder affects the two transitions differently. For x=0.50, only one peak corresponding to the SG2 transition is seen clearly. There is, however, an inflection point around 51K which could possibly be linked with the SG1 transition. The SG1 transition temperatures for various compositions, including x= 0.40 and 0.50, also show $T_c \sim (x-x_c)^n$ type dependence with $x_c$ =0.55±0.01 but with an exponent n =0.49±0.07. This exponent (n~ ½) is reminiscent of a quantum phase transition [82,83] and the possibility of the existence of a quantum critical point corresponding to the percolation threshold $x_c$=0.55 for the SG1 transition needs to be investigated carefully in a future work. In contrast to the SG1 transition, the SG2 transition temperature shows weak composition dependence upto about x= 0.40 but significant decrease is seen for x=0.50. From the least squares fit to the observed $T_{f2}$ values using $(x-x_c)^n$ type dependence, the critical composition limit for this transition is also found to be close to $x_c$=0.55 but with an exponent n =0.08.

We have also examined the composition dependence of the intermediate transition that occurs between the SG1 and SG2 transitions which is known to be driven by electromagnons in pure BF, using ZFC M(T) plot below room temperature shown in *Fig.20* for four different *compositions of BF-xBT. The M(T) shows a peak corresponding to the SG1 transition whereas the SG2 transition is signalled by a decrease in the magnetisation value at low temperatures. As a result of dilution of the magnetic sublattice due to disorder, magnetization decreases and the peak corresponding to the SG1 transition becomes less prominent for x=0.40. The electromagnon transition is signalled by a kink (for x≤ 0.30) or a dip (x=0.40) at the foothill of the SG1 peak. The corresponding transition temperature shows a rather weak composition dependence upto x=0.30. The composition dependence of this transition temperature ($T_c$) was also fitted to $(x-x_c)^n$ type function which gave us n= 0.33±0.06 and $x_c$ =0.55±0.02. The phase diagram presented here clearly shows that the SG1 and SG2 transitions are intervened by a third transition supposedly driven by electromagnons for all the compositions with x < $x_c$, a situation not envisaged in the existing theories of a succession of two spin glass transitions in Heisenberg systems [23-26].*

**I. Concluding remarks:**

*We have presented evidence for two spin glass transitions in the BF-xBT system using a series of bulk measurements revealing history dependent effect, critical slowing down of the spin dynamics due to ergodicity breaking, existence of A-T and G-T lines due to freezing of the longitudinal and transverse components of the spins and stretched exponential decay of the thermoremanent magnetization. Using neutron and x-ray diffraction measurements, which provide evidence on microscopic scales, we have shown that the two spin glass transitions are not only intrinsic to the BF-xBT system but also occur on the same magnetic sublattice in coexistence with the long range ordered antiferromagnetic phase. Our results show for the first time that the spontaneous polarization ($P_s$) and unit cell volume (V) show significant variation across the SG1 and SG2 transitions confirming the presence of magnetoelectric and*

*magnetoelastic couplings, respectively. These couplings and the possibly the presence of electromagnons constitute unique features of a multiferroic spin glass systems that distinguish them from the conventional spin glass systems. While the existence of the A-T and G-T lines confirm that the SG1 and SG2 transitions result from the freezing of the longitudinal and transverse components of spins as predicted theoretically for Heisenberg systems with small single ion anisotropy (D), there are a few unexplained aspects of our observations. First and foremost is whether the smeared SG1 transition could have a structural origin, rather than magnetic. Although the SG1 transition is not found to be linked with any change in the space group symmetry, the occurrence of isostructural phase transition (ISPT) has been confirmed by us which indicates spin-phonon coupling. Secondly, the temperatures for the two spin glass transitions are far too apart whereas the difference between the two-successive spin-glass transitions in conventional spin glasses is rather modest (<50K). Thirdly, the two spin glass transitions are not successive as there is another transition, possibly driven by electromagnons, in between the two spin glass transitions. Any plausible theory of spin glass transitions in a multiferroic system requires consideration of magnetoelastic and magnetoelectric couplings as well as electromagnons, if present. The mechanism of spin-phonon coupling (electronic or otherwise) needs to be investigated for each multiferroic system since it differs from compound to compound [80]. We hope that our results would stimulate future work to consider the effect of these couplings and electromagnons in the mean field theories as well as Monte Carlo simulation studies of SG transitions in insulating magnetoelectric multiferroics like $BiFeO_3$.*

**Acknowledgement:** We thank Prof. Chalapathi N. V. Rao, Department of Geology, Banaras Hindu University, Varanasi-221005 for the EPMA measurements. DP acknowledges financial support from Science and Engineering Research Board (SERB) of India through the award of J C Bose National Fellowship and UGC-DAE CSR, Indore through a project.

**Figure Captions:**

**Fig. 1.** ZFC DC magnetization vs temperature plot for an applied field of 500 Oe. Insets depict (a) the temperature dependence of DC magnetization under ZFC and FC conditions and (b) Curie-Weiss plot for BF-0.20BT above $T_N$.

**Fig. 2.** The M-H hysteresis loop at 300K for BF-0.20BT.

**Fig. 3.** Neutron powder diffraction pattern at room temperature. Arrow marks the antiferromagnetic peak. All the indices are written with respect to a doubled pseudocubic cell.

**Fig. 4.** The variation of specific heat capacity with temperature for BF-0.20BT. Inset (a) is a magnified view around SG1 transition depicting an anomaly.

**Fig. 5.** Variation of $\chi'(\omega, T)$ and $\chi''(\omega, T)$ in the temperature range 2-300K at various frequencies [47.3 Hz (►), 97.3 Hz (◄), 197.3 Hz (▼), 297.3 Hz (▲), 397.3 Hz (●), 497.3 Hz (■)]. Insets (i) and (ii) depict $\chi'(\omega, T)$ on a zoomed scale for SG 1 and SG 2, respectively.

**Fig. 6.** $\ln\tau$ versus $1/T$ plot for (a) SG1 and (b) SG 2 transitions. Solid line is the least squares fit for Vogel-Fulcher law.

**Fig. 7.** $\ln\tau$ versus $\ln(T-T_{SG}/T_{SG})$ plot for (a) SG1 and (b) SG 2 transitions. Solid line shows the least squares fit for power law.

**Fig. 8.** ZFC DC magnetization vs temperature plots of BF-0.20BT measured at different applied fields. Insets depict the magnified view around SG2 transition.

**Fig. 9.** (a) de Almeida-Thouless (A-T) line for SG1 transition and (b) Gabay-Toulouse (G-T) line for SG2 transition.

**Fig. 10.** Variation of thermoremanent remnant magnetization (M (t)) with time at (a) 200 K and (b)10 K for BF-0.20BT.

**Fig. 11.** The evolution of x-ray powder diffraction profiles of the $(222)_p$, $(400)_p$ and $(440)_p$ reflections of BF-0.20BT with temperature showing absence of any structural phase transition.

**Fig. 12.** Observed (filled circles), calculated (continuous line), and difference (bottom line) profiles obtained from the Rietveld refinement using SXRD data at (a) 220K (b) 240K and (c) 260K using R3c space group for BF-0.20BT. The vertical tick marks above the difference profile represent the Bragg peak positions.

**Fig. 13.** Variation of unit cell volume with temperature: XRD (▲) and NPD (●) data. Solid line (━) is fit for Debye Grüneisen equation $T_{SG1}$. Inset (a) shows the zoomed view around 140K and SG2 transitions. Inset (b) depicts the variation of volume strain ($\Delta V/V$) against square of magnetization ($M_S^2$) obtained by M-H loop.

**Fig. 14.** The evolution of the neutron powder diffraction patterns with temperature in the limited $2\theta = 15°-57°$ range. The first peak is due to AFM ordering. The Miller indices are written with respect to a doubled pseudocubic cell.

**Fig. 15.** Temperature dependent variation of the integrated intensity of the AFM peak (111) (The miller indices are with respect to a doubled pseudocubic cell). Solid line is fit for Brillouin function.

**Fig. 16.** Temperature dependence of the fractional z coordinates of (a) Bi/Ba and (b) Fe/Ti. The x and y coordinates of O are shown in (c) and (d). All these coordinates were obtained from the Rietveld refinements using neutron powder diffraction data.

**Fig. 17.** Temperature dependent variation of the spontaneous polarization calculated from the positional coordinates.

**Fig. 18.** (a) The variation of ZFC magnetization with temperature measured at a field of 500 Oe for various compositions in the range $0.10 \leq x \leq 0.60$. (b) shows first derivative of M (dM/dT) with respect to temperature for these compositions.

**Fig. 19.** Left panel shows the variation of $\chi'(\omega, T)$ of BF-xBT with temperature at 497.3 Hz frequency for various compositions in the range $0.10 \leq x \leq 0.60$. Right panel (a-c) as well as panel (d) depict the zoomed view around the SG1 transition.

**Fig.20.** The variation of ZFC magnetization of BF-xBT with temperature below 300K measured at field of 500 Oe for compositions (a) x= 0.10, (b) x= 0.20, (c) x= 0.30 and (d) x= 0.40

**Fig.21.** Phase diagram of BF-xBT. PM: Paramagnetic, SG: Spin glass, AFM: Antiferromagnetic, EM: Electromagnon. The SG2 transition temperatures (see the inset) shows the weakest composition dependence. The dotted lines through the data points depict the least squares fit for $T_c \sim (x-x_c)^n$ type dependence with $x_c = 0.55$ giving n = 0.30, 0.49, 0.33 and 0.08 for the AFM, SG1, electromagnon driven and SG2 transitions, respectively. The exponent n ~ ½ indicates the possibility of a quantum critical point at $x_c \sim 0.55$.

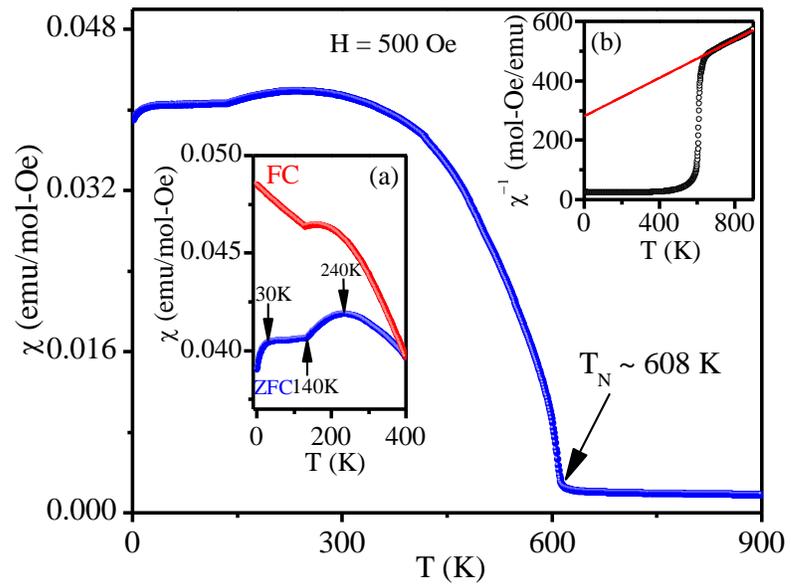

**Fig. 1**

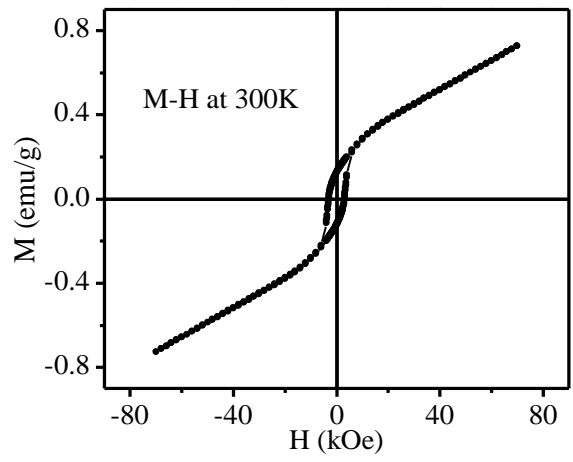

**Fig. 2**

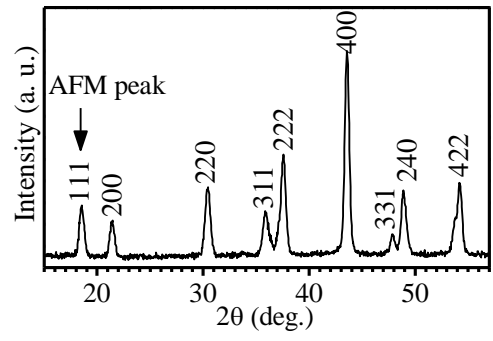

**Fig. 3**

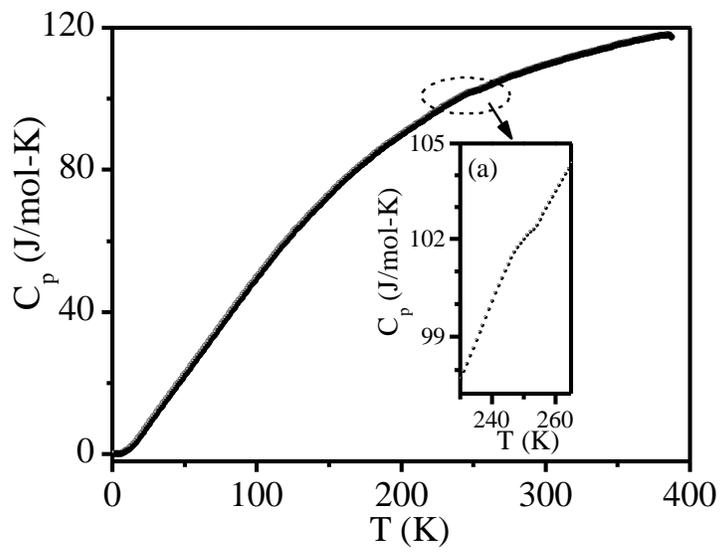

**Fig. 4**

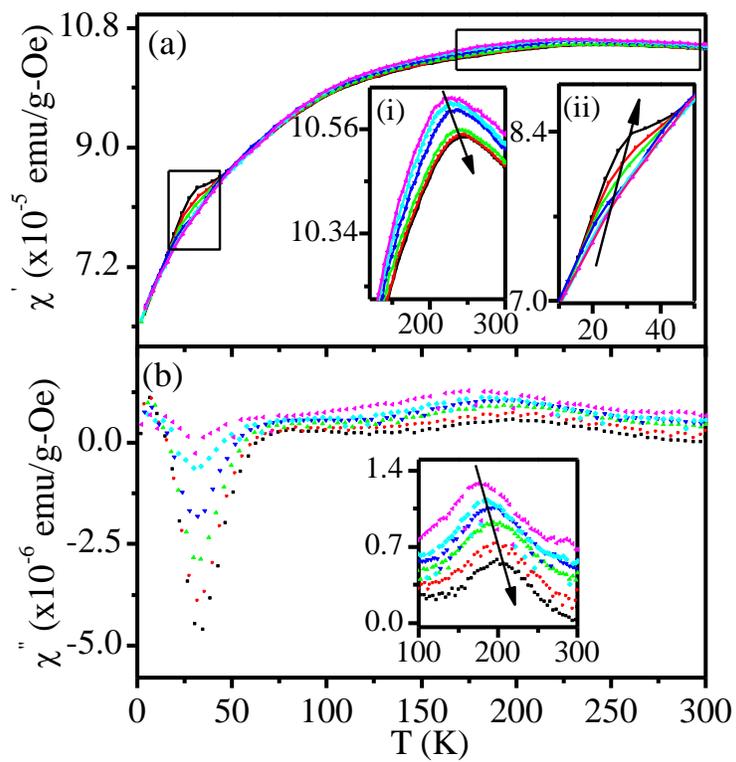

**Fig. 5**

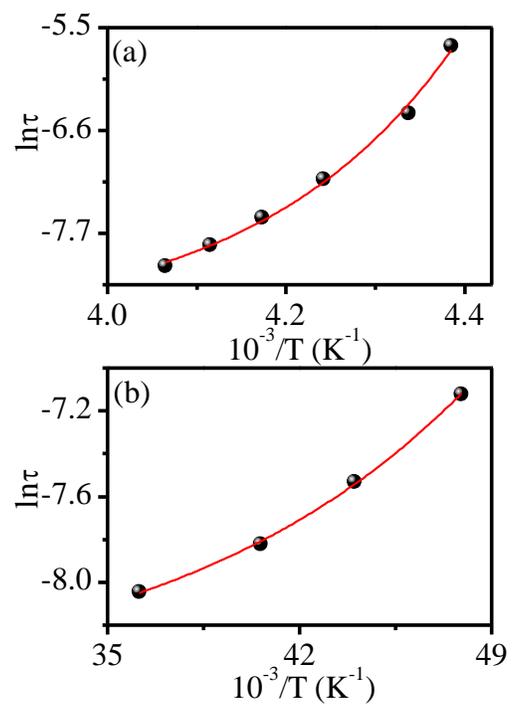

**Fig. 6**

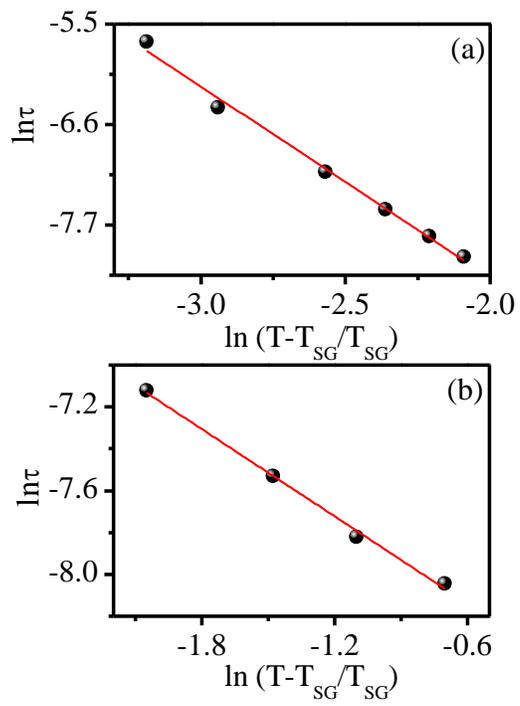

**Fig. 7**

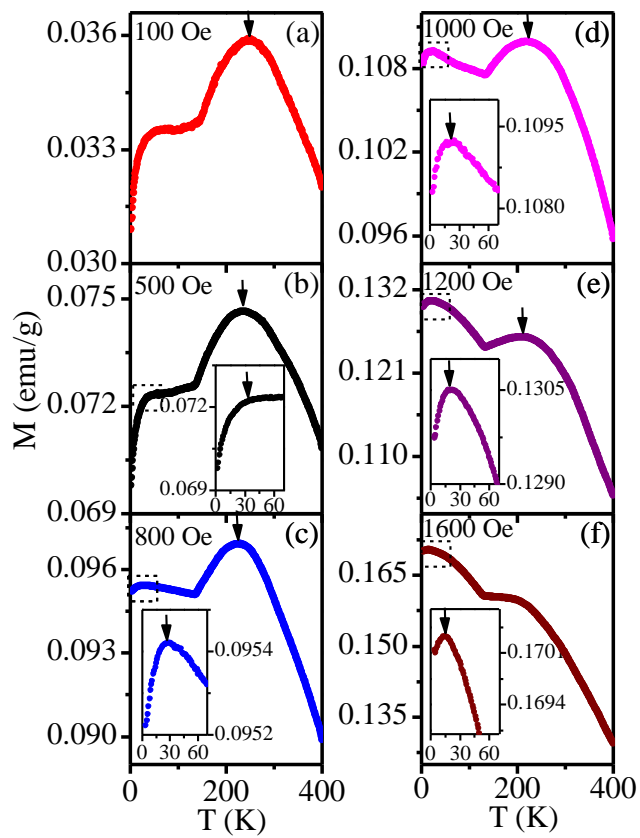

**Fig. 8**

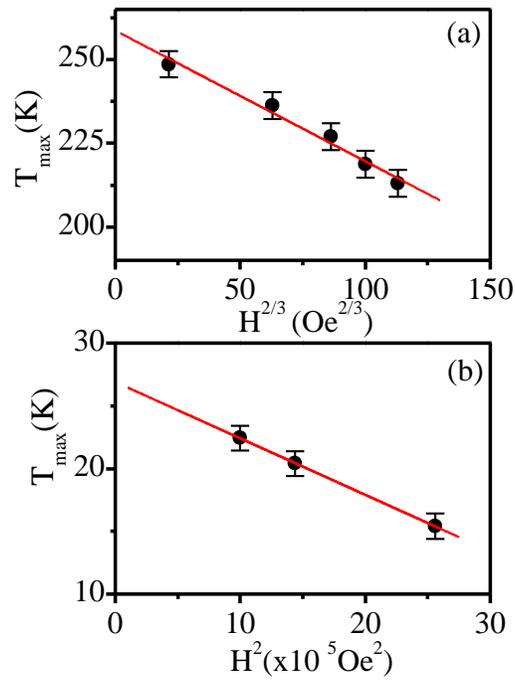

**Fig. 9**

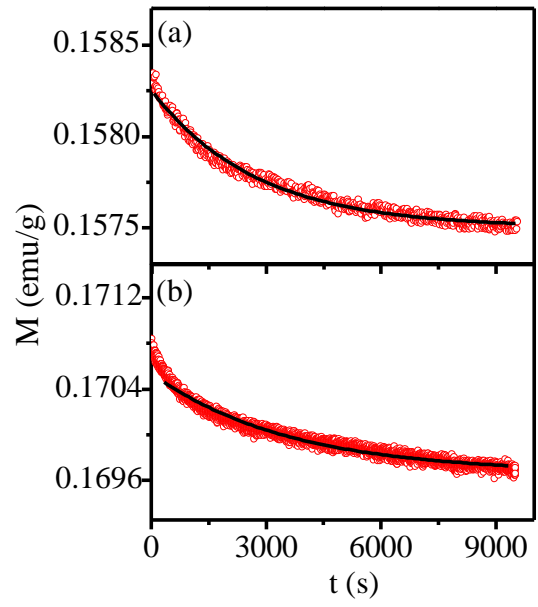

**Fig. 10**

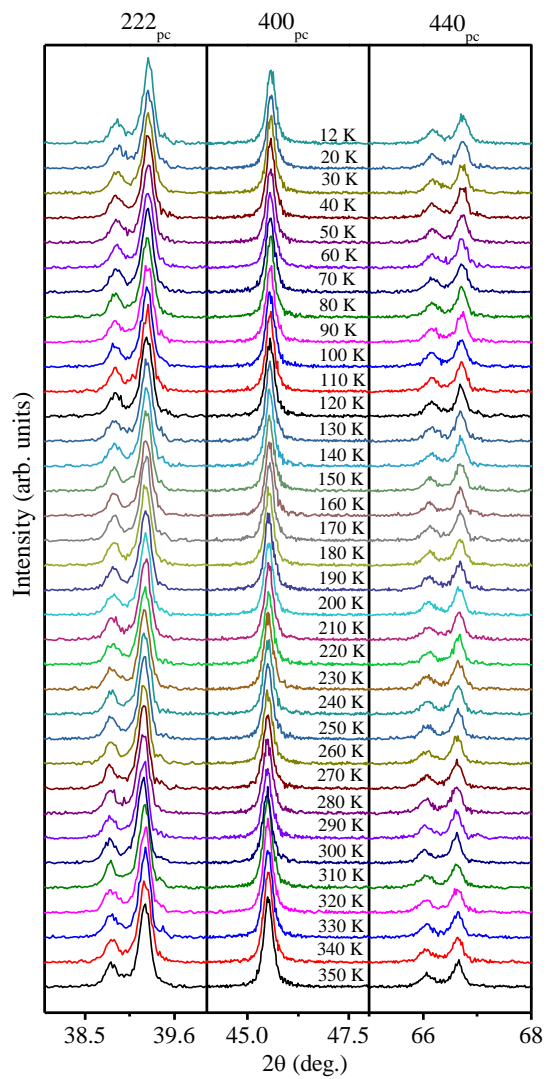

**Fig. 11**

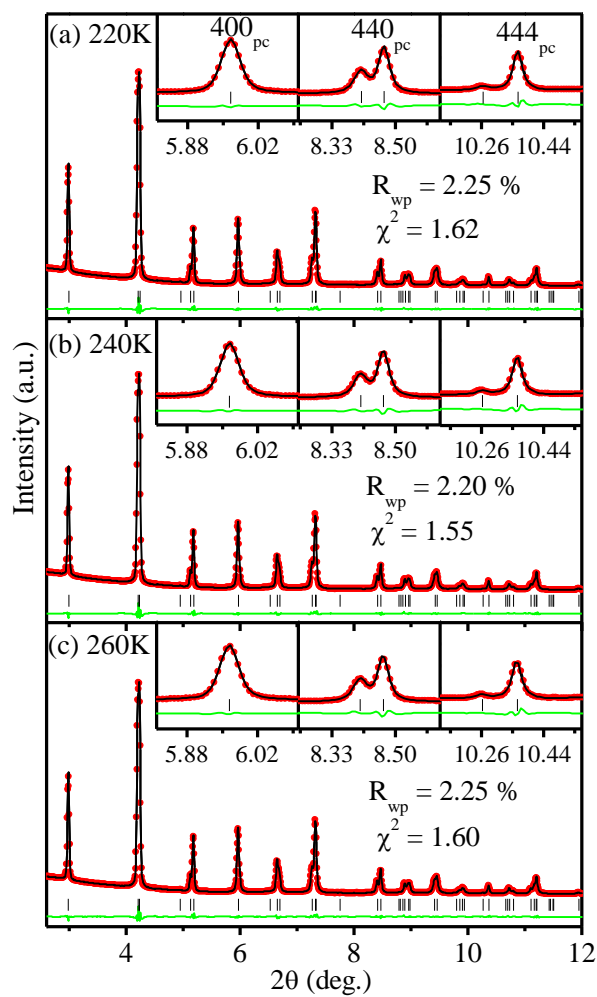

**Fig. 12**

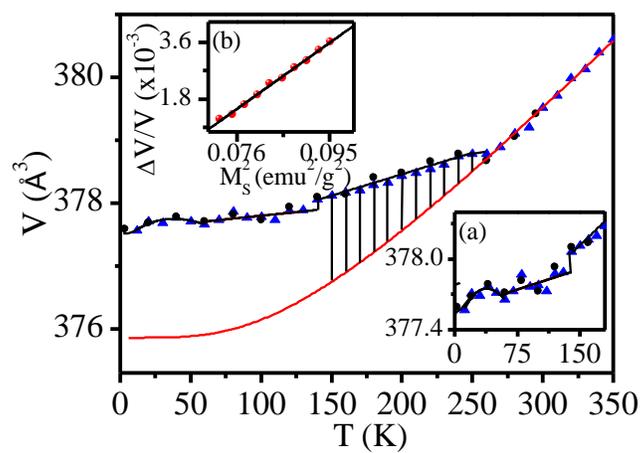

**Fig. 13**

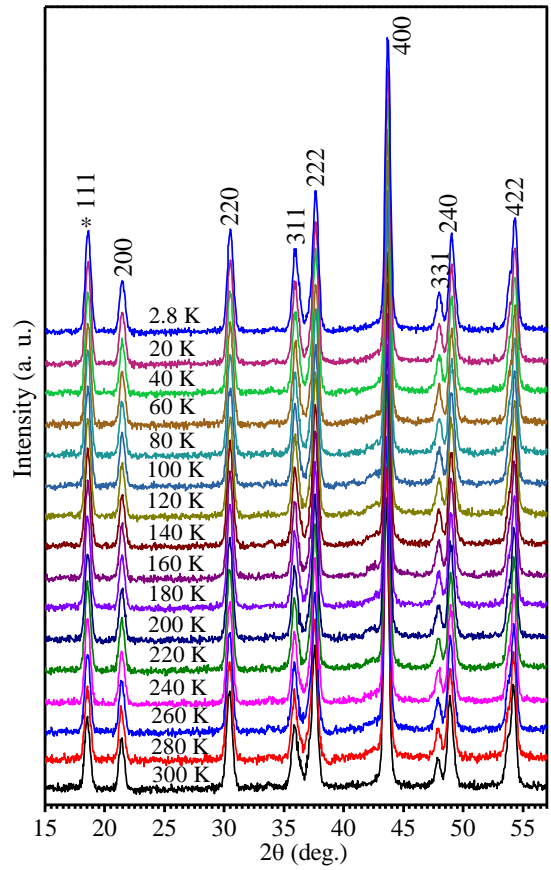

**Fig. 14**

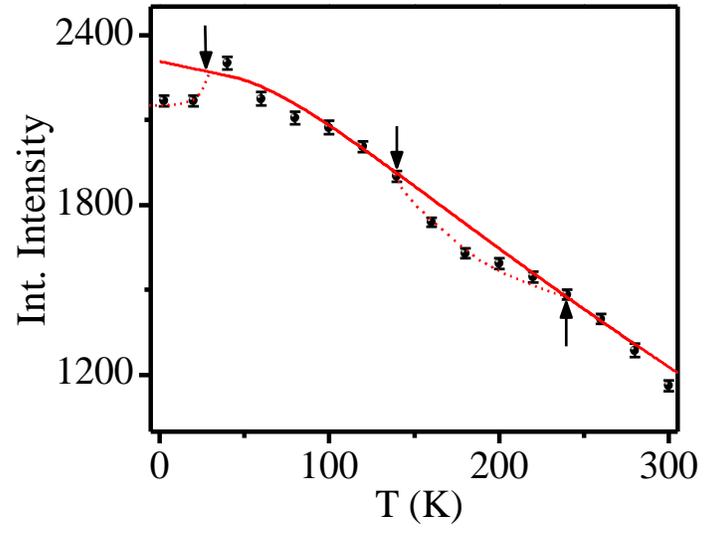

**Fig. 15**

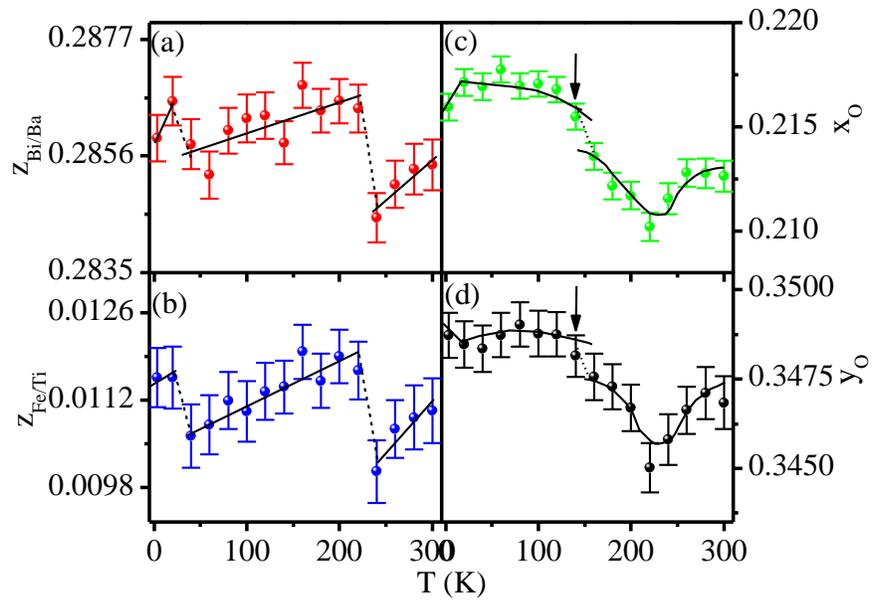

**Fig. 16**

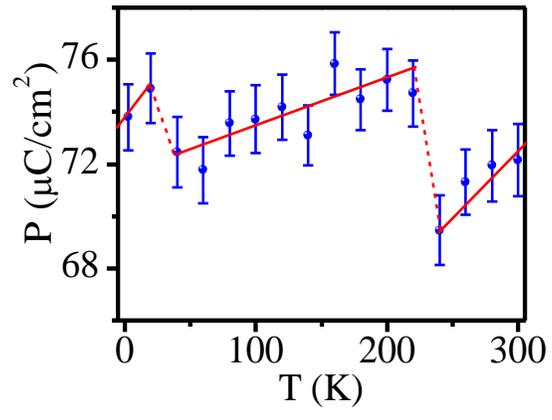

**Fig. 17**

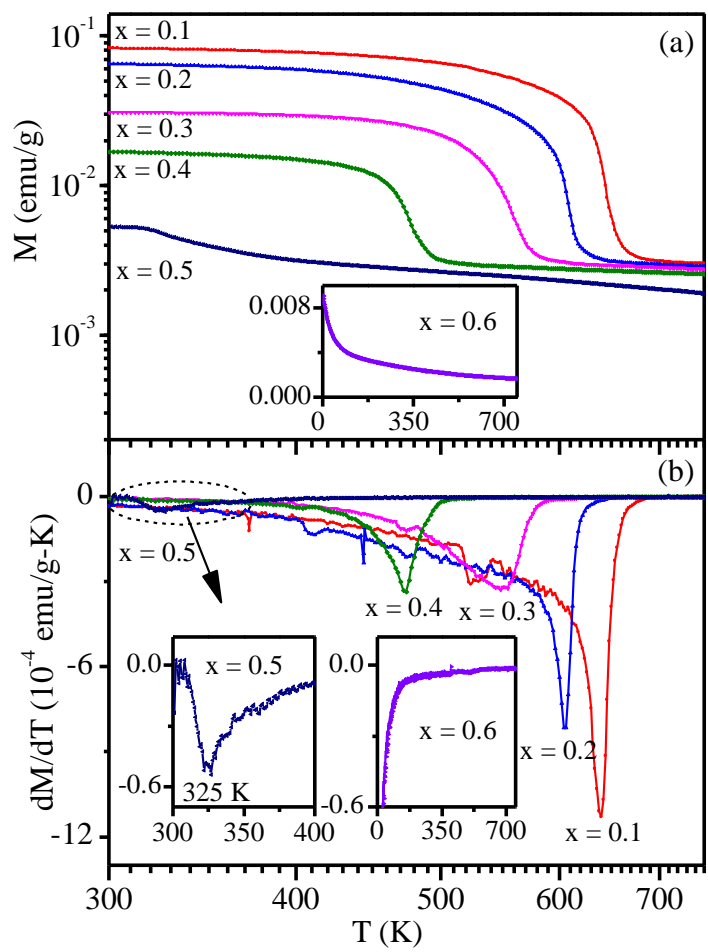

**Fig. 18**

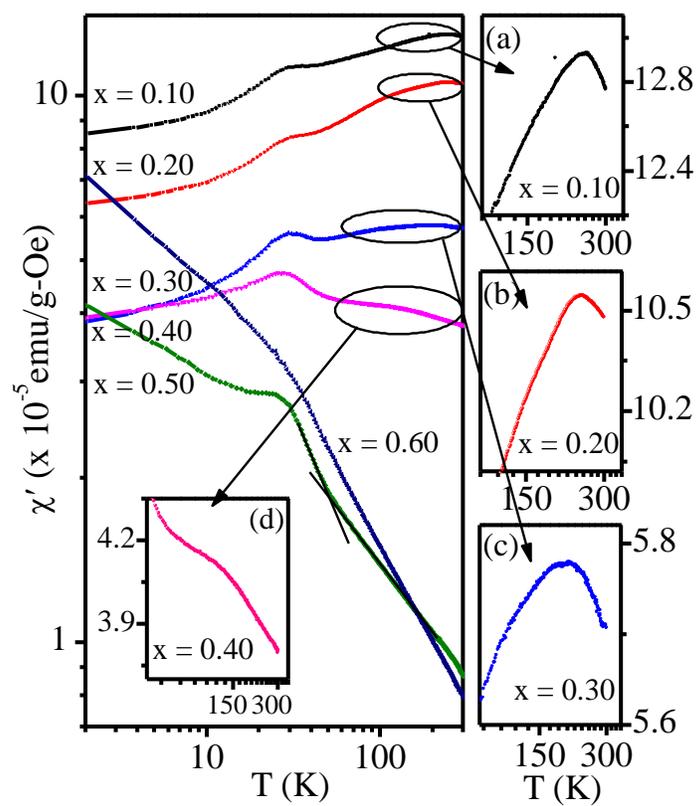

**Fig. 19**

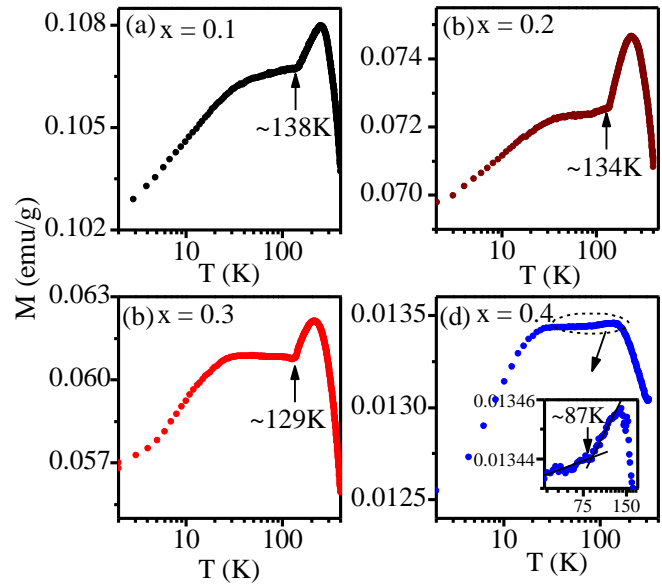

**Fig. 20**

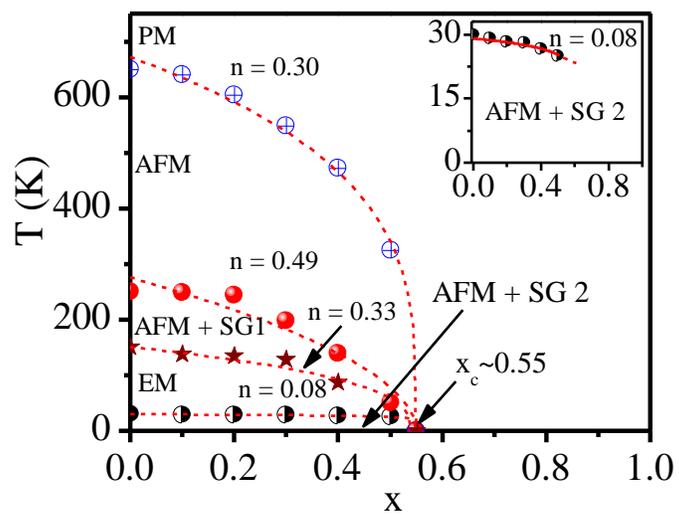

**Fig. 21**

# Supplemental Information

**S1: Chemical composition and phase purity analysis of BF-0.2BT:**

The results of the quantitative analysis of the chemical composition of BF-0.2BT sample using EPMA, averaged over 10 different regions, are given in Table S1 along with the standard deviation. *It is evident from the table that the values obtained by EPMA analysis are close to the nominal composition within the standard deviation. This confirms excellent sample quality.*

The synchrotron x-ray diffraction (SXRD) pattern of the sintered powder of BF-0.2BT at room temperature is shown in Fig.S1. *It is evident from the figure that all the peaks in the SXRD patterns of the sintered powder of BF-0.2BT could be indexed with the pure perovskite phase and no trace of any impurity phase is observed.*

Table S1: Compositional analysis of BF-0.2BT sample using EPMA

| Average Chemical Composition in Weight % | | |
|---|---|---|
| Element | Expected | Average |
| Bi | 56.3 | $56.2 \pm 0.5$ |
| Fe | 15.1 | $15.0 \pm 0.2$ |
| Ba | 9.2 | $8.9 \pm 0.2$ |
| Ti | 3.2 | $3.1 \pm 0.05$ |
| Mn | 0.3 | $0.22 \pm 0.04$ |
| O | 16.2 | $15.2 \pm 0.5$ |

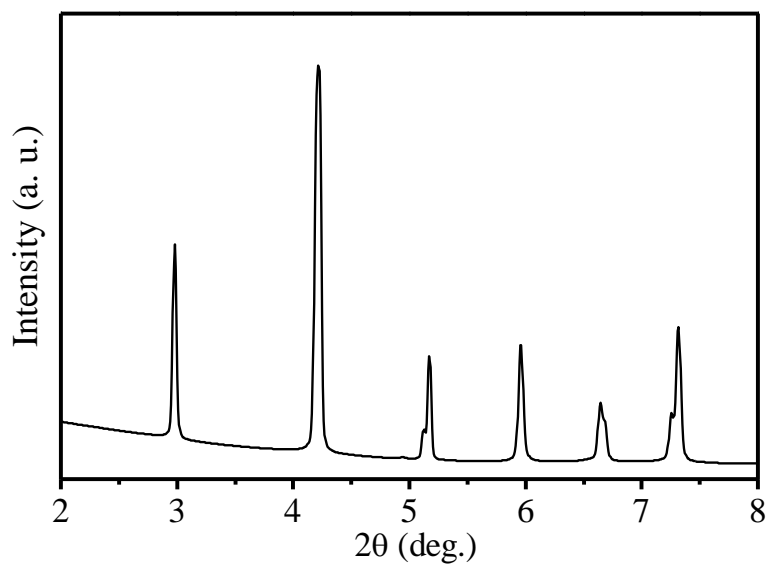

**Fig. S1.** The SXRD pattern of the sintered BF-0.2BT powder at room temperature.

**S2: Rietveld refinement at room temperature using synchrotron x-ray diffraction pattern:**

The asymmetric unit of rhombohedral structure with R3c space group consists of three ions ($Bi^{3+}$/ $Ba^{2+}$, $Fe^{3+}$/ $Ti^{4+}$ and $O^{2-}$) in which, $Bi^{3+}$/ $Ba^{2+}$ and $Ti^{4+}$/$Fe^{3+}$ ions occupy the 6(a) Wyckoff site at (0, 0, z) while $O^{2-}$ ions at the 18(b) sites at (x, y, z) in the hexagonal unit cell. Following Megaw and Darlington notation [1], the positional coordinates of atoms in the asymmetric unit cell can be written as $Bi^{3+}$/$Ba^{2+}$ (0,0,1/4+s), $Fe^{3+}$/$Ti^{4+}$ (0,0,t), $O^{2-}$ (1/6-2e-2d,1/3-4d,1/12). The parameters s and t describe the displacement of cations along $[111]_{pc}$ axis, whereas d and e represent the octahedral distortion and octahedral tilt angle $\omega = \tan^{-1}(4e\sqrt{3})$ along $[111]_{pc}$ axis, respectively [1]. In the refinement process the background was modeled with linear interpolation and the peak shape was modeled using pseudo-Voigt function. Occupancy of all the ions were fixed at the nominal composition in the refinements. Zero correction, scale factor, background, lattice parameters, half width parameters (u, v and w), positional coordinates and thermal parameters were varied during the refinement. Fig.S2 depicts the observed, calculated and difference profiles obtained after the Rietveld analysis of the SXRD pattern for BF-0.2BT using R3c space group. The observed (filled-circles) and calculated (continuous line) profiles are in excellent agreement, as can be seen from the difference (bottom line) profile given in Fig.S2. This confirms that all the peaks in the SXRD pattern of the BF-0.2BT samples are indexed with single phase of rhombohedral structure with R3c space group. The refined structural parameters given in Table S2 are in good agreement with those reported in literature [2, 3].

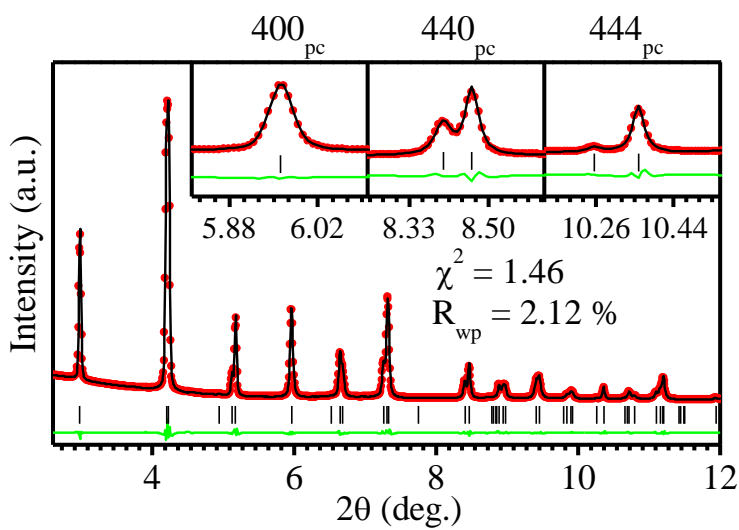

**Fig.S2:** Observed (filled circles), calculated (continuous line), and difference (bottom line) profiles obtained from the Rietveld refinement of SXRD data at room temperature using R3c space group for BF-0.2BT. The vertical tick marks above the difference profile represent the Bragg peak positions.

**S3: Low temperature x-ray diffraction (XRD) studies:**

We have verified the absence of structural phase transition in BF-0.2BT below room temperature by x-ray diffraction patterns (XRD) using Rietveld technique. The asymmetric unit of rhombohedral structure with R3c space group is already given in section S2. The refinement converged successfully after a few cycles at all temperatures. The excellent fits confirm the R3c space group for BF-0.2BT at all temperatures. Fig.12 depicts the observed, calculated and difference profiles obtained after the Rietveld analysis of the XRD patterns at selected temperatures 300K, 200K, 100K and 12K, respectively, for BF-0.2BT using R3c space group. Thus, our Rietveld refinements confirm that there is no structural phase transition down to l2K.

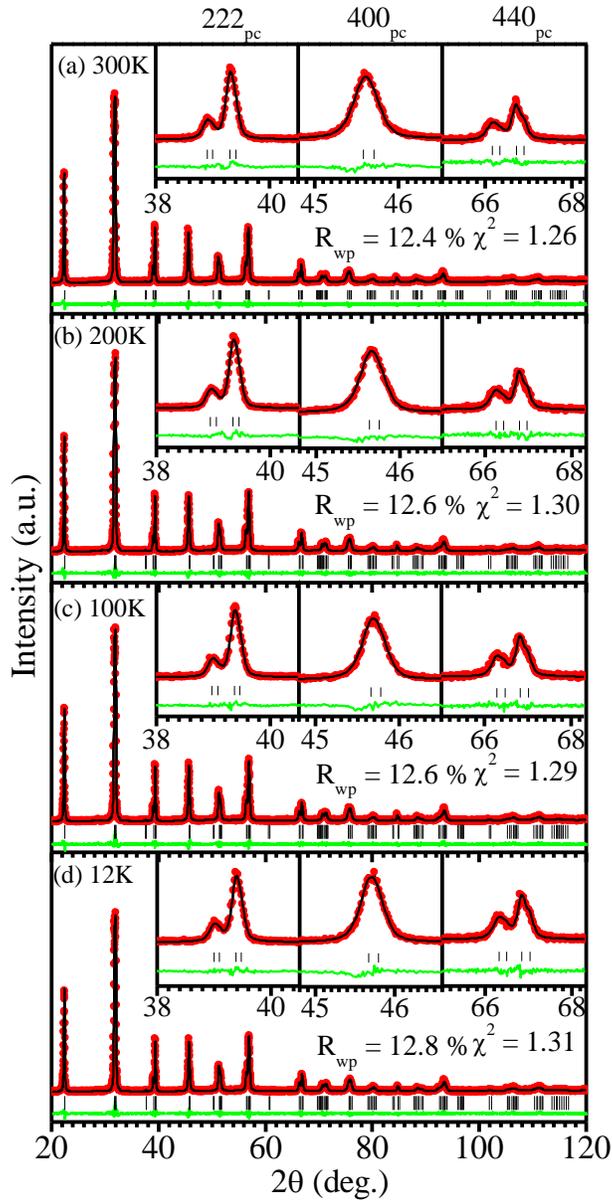

Fig. S3. Observed (filled circles), calculated (continuous line), and difference (bottom line) profiles obtained from Rietveld refinement using R3c space group at (a) 300K (b) 200K (c)100K and (d) 12K. The vertical tick marks correspond to the position of all allowed Bragg reflections.

**S4: Neutron powder diffraction (NPD) studies:**

In this section, we present the details of the Rietveld analysis of the NPD patterns. The asymmetric unit of rhombohedral structure with R3c space group is already given in section S2. All the nuclear structure peaks are well indexed with respect to unit cell of the R3c space group except the magnetic peaks. *No evidence for any magnetic impurity phase was found in the neutron data.* The magnetic peaks are indexed by considering additional phase in the nuclear structure refinement of neutron powder diffraction (NPD) data. Following Singh et al. [2, 3], all the magnetic peaks were well indexed with propagation vector $\vec{k}$ = (0,0,0). The initial input parameters for Rietveld refinement of nuclear structure were taken from the Rietveld refinement using SXRD data. Both the nuclear and magnetic structures were refined, and the refinement converged successfully after a few cycles. The observed (filled-circles) and calculated (continuous line) profiles show excellent fits at all temperatures and some selected Rietveld refined profiles (at 300 K, 200K, 100K and 2.8K) are shown in Fig. S4 (a) (b) (c) and (d), respectively. The refined lattice parameters, positional coordinates, thermal parameters, and magnetic moment are listed in Table S2 are in good agreement with those reported in literature [2, 3]. Our Rietveld refinement results also confirm that the nuclear structure with R3c space group does not change down to the lowest temperature as a result of the magnetic transitions.

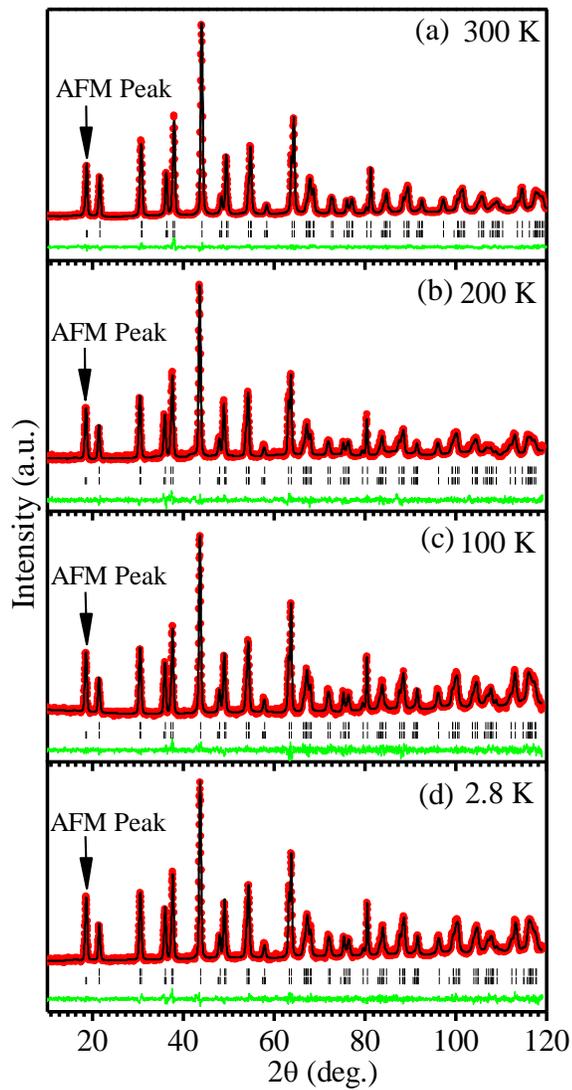

Fig. S4. Observed (filled circles), calculated (continuous line), and difference (bottom line) profiles obtained from Rietveld refinement using R3c space group at (a) 300K (b) 200K (c) 100K and (d) 2.8K. Arrow indicates AFM peak. The vertical tick marks correspond to the position of all allowed Bragg reflections for the nuclear (top) and magnetic (bottom) reflections.

**Table S2**: Refined nuclear and magnetic structural parameters and agreement factors using SXRD data at 300K and NPD data at 300K, 200K, 100K and 2.8K with R3c space group.

| Parameters | SXRD 300K | NPD 300 K | NPD 200 K | NPD 100 K | NPD 2.8 K |
|---|---|---|---|---|---|
| $a_{hex}$ (Å) | 5.6139 (2) | 5.6132 (5) | 5.6084 (5) | 5.6052 (5) | 5.6051 (4) |
| $c_{hex}$ (Å) | 3.9066 (5) | 13.9078 (2) | 13.8939 (1) | 13.8825 (1) | 13.8813 (1) |
| $v_{hex}$ (Å) | 379.48 (5) | 379.42 (6) | 378.48 (6) | 377.73 (6) | 377.59 (5) |
| α, β, γ | α=β=90°, γ=120° | α=β=90°, γ=120° | α=β=90°, γ=120° | α=β=90°, γ=120° | α=β=90°, γ=120° |
| Bi/Ba (z) | 0.2867 (6) | 0.2854 (5) | 0.2865 (4) | 0.2862 (4) | 0.2859 (4) |
| Fe/Ti (z) | 0.0121 (5) | 0.0110 (6) | 0.0119 (4) | 0.0110 (5) | 0.0115 (4) |
| O (x) | 0.2104 (5) | 0.2116 (8) | 0.2117 (6) | 0.2171 (6) | 0.2159 (6) |
| O (y) | 0.3461 (7) | 0.3468 (4) | 0.3466 (7) | 0.3488 (6) | 0.3487 (6) |
| O (z) | 1/12 | 1/12 | 1/12 | 1/12 | 1/12 |
| $\beta_{Bi/Ba}$ (Å$^2$) | $\beta_{11}=\beta_{22}=2\beta_{12}$ = 0.0431 (3) <br> $\beta_{33}$ = 0.0039 (5) | $\beta_{11}=\beta_{22}=2\beta_{12}$ = 0.0363 (2) <br> $\beta_{33}$ = 0.0031 (3) | $\beta_{11}=\beta_{22}=2\beta_{12}$ = 0.0313 (2) <br> $\beta_{33}$ = 0.0029 (4) | $\beta_{11}=\beta_{22}=2\beta_{12}$ = 0.0250 (2) <br> $\beta_{33}$ = 0.0026 (3) | $\beta_{11}=\beta_{22}=2\beta_{12}$ = 0.031 (2) <br> $\beta_{33}$ = 0.0024 (2) |
| $\beta_{Fe/Ti}$ (Å$^2$) | 1.36 (5) | 1.33 (9) | 1.30 (5) | 1.27 (9) | 1.27 (8) |
| $\beta_O$ (Å$^2$) | $\beta_{11}$= 0.064 (8) <br> $\beta_{22}$ = 0.022 (3) <br> $\beta_{33}$ = 0.004 (3) <br> $\beta_{12}$ = 0.029 (4) <br> $\beta_{13}$ = 0.004 (8) <br> $\beta_{23}$ = 0.008 (3) | $\beta_{11}$= 0.057 (3) <br> $\beta_{22}$ = 0.019 (1) <br> $\beta_{33}$ = 0.002 (3) <br> $\beta_{12}$ = 0.025 (2) <br> $\beta_{13}$ = 0.003 (9) <br> $\beta_{23}$ = 0.006 (4) | $\beta_{11}$= 0.048 (5) <br> $\beta_{22}$ = 0.011 (3) <br> $\beta_{33}$ = 0.002 (3) <br> $\beta_{12}$ = 0.019 (7) <br> $\beta_{13}$ = 0.003 (6) <br> $\beta_{23}$ = 0.006 (5) | $\beta_{11}$= 0.039 (3) <br> $\beta_{22}$ = 0.004 (13) <br> $\beta_{33}$ = 0.003 (2) <br> $\beta_{12}$ = 0.014 (2) <br> $\beta_{13}$ = 0.002 (7) <br> $\beta_{23}$ = 0.005 (3) | $\beta_{11}$= 0.046 (3) <br> $\beta_{22}$ = 0.008 (14) <br> $\beta_{33}$ = 0.003 (2) <br> $\beta_{12}$ = 0.017 (1) <br> $\beta_{13}$ = 0.004 (7) <br> $\beta_{23}$ = 0.006 (4) |
| $\mu_{Fe}$ ($\mu_B$) | - | 3.25 (8) | 3.55 (7) | 3.75 (8) | 3.82 (7) |
| $R_{wp}$ (%) | 2.12 | 9.15 | 9.89 | 9.87 | 9.03 |
| $R_{mag}$ (%) | - | 4.63 | 4.21 | 3.39 | 3.36 |
| $\chi^2$ | 1.46 | 7.58 | 8.35 | 9.80 | 8.38 |

**Supplementary References:**